\documentclass{ws-procs9x6}

\newcommand{\refeq}[1]{(\ref{#1})}
\def\etal {{\it et al.}}

\begin{document}

\begin{flushright}IUHET 548, July 2010\end{flushright}

\title{LORENTZ AND CPT VIOLATION IN NEUTRINO OSCILLATIONS}

\author{JORGE S.\ D\'IAZ }

\address{Physics Department, Indiana University,\\
Bloomington, IN 47405, USA\\
E-mail: jsdiazpo@indiana.edu}

\begin{abstract}
Neutrino oscillations in the presence of Lorentz violation can present novel observable signals in both long- and short-baseline experiments. In this talk we describe the theory and its different regimes depending on properties of the experiments. CPT violation, its systematic search and possible connections to latest results are also presented.
\end{abstract}

\bodymatter

\section{Introduction}
\label{JSD_Sec1}

Neutrino oscillations have become a powerful method to test our understanding of the Standard Model (SM) of elementary particles. Indeed, the very fact that neutrinos change flavor as they propagate cannot be explained by the SM. The conventional description of neutrino oscillations requires these particles to have tiny masses. Most of the current neutrino data can be accommodated within this description; nevertheless, during recent years several results suggest that our understanding of neutrino oscillations might be incomplete. In the search for new physics, we can take advantage of the interferometric nature of neutrino oscillations, which opens a window to the study of small-scale effects challenging to address directly. One of the promising suppressed effects that could be observed at low energies is the breaking of Lorentz symmetry. In the study of candidate quantum descriptions of gravity, it has been proved that Lorentz symmetry breaking might arise naturally at the Planck scale.\cite{KS1989} The possible effects produced by Lorentz violation would be suppressed signals at low energies that would be observed as deviations from the Lorentz-invariant description of a given phenomenon. Observable signals of Lorentz violation can be described using effective field theory,\cite{KP1995} which is independent of the underlying theory and contains all possible terms that can be added to the SM that break Lorentz symmetry. The Standard-Model Extension (SME) is such a general framework.\cite{CK_SME} In flat spacetime, the SME is constructed by adding all possible coordinate-independent terms to the SM lagrangian, which consist of SM operators properly contracted with controlling coefficients for Lorentz violation. Since CPT violation implies the breaking of Lorentz symmetry,\cite{greenberg} general CPT violation is also included within the SME. We will restrict our attention to the minimal SME (mSME), which involves renormalizable terms only.

The SME is a framework used worldwide in searches for Lorentz violation. From the theory point of view, the SME is an effective field theory that preserves the SU(3)$\times$SU(2)$\times$U(1) gauge structure of the SM, including the SU(2)$\times$U(1) symmetry breaking, energy-momentum conservation, hermiticity, positivity of energy, and anomaly cancellation. Experimentally, the SME is a robust framework to search for Lorentz violation that allows us to relate results across different disciplines in a physically meaningful way. Additionally, the SME can be used to estimate the observable effects to look for in a determined experiment.

\section{Lorentz-violating neutrino oscillations}
\label{JSD_Sec2}

The neutrino sector of the mSME describes the behavior of three active left-handed neutrinos by the effective hamiltonian\cite{KM2004a}

\begin{equation}\label{JSD.eq01}
(h_\text{eff})_{ab} =\frac{m^2_{ab}}{2E}+\frac{1}{E}\big[(a_L)^\alpha p_\alpha-(c_L)^{\alpha\beta} p_\alpha p_\beta\big]_{ab},
\end{equation}
where the first term is the conventional Lorentz-invariant mass matrix, the second term controls both Lorentz and CPT violation, and the third term controls Lorentz violation only. Here we have not included the term $E\delta_{ab}$ because it does not contribute to oscillations; nonetheless, for stability and causality of the fundamental theory this term can be relevant.\cite{KL2001}
The subscripts are flavor indices $a,b=e,\mu,\tau$. The corresponding hamiltonian for right-handed antineutrinos can be obtained by replacing $m^2_{ab}\rightarrow(m^2_{ab})^*$, $(a_L)^\alpha_{ab}\rightarrow-(a_L)^{\alpha*}_{ab}$, and $(c_L)^{\alpha\beta}_{ab}\rightarrow(c_L)^{\alpha\beta*}_{ab}$ in Eq. \refeq{JSD.eq01}. The oscillatory behavior of the oscillation probability as a function of the energy arises from the term $\sin^2(\Delta_{a'b'}L/2)$, where $\Delta_{a'b'}$ is the difference of the eigenvalues of the effective hamiltonian. The first term in Eq. \refeq{JSD.eq01} leads to the standard $L/E$ dependence, whereas the Lorentz-violating terms introduce oscillation phases that are constant $(a_L)^\alpha_{ab}L$ and that grow with the energy $(c_L)^{\alpha\beta}_{ab}LE$. This is one of the key signals of Lorentz violation because a Lorentz invariant description requires the oscillation phase to decrease with the energy.
It is important to mention that the hamiltonian could have a large term at high energies triggering a Lorentz-violating seesaw mechanism, in which case the coefficients $a_L$ and $c_L$ generate an oscillation phase that takes the conventional form $L/E$. This means that even in the absence of mass, neutrinos could present an oscillatory behavior that looks like a mass. This mechanism appears in both the bicycle and the tandem models.\cite{KM2004b,tandem} Nonetheless, in this talk we will focus on another key signal of Lorentz violation. In the effective hamiltonian \refeq{JSD.eq01}, the coefficients for Lorentz violation $a_L$ and $c_L$ are coupled to the four-momentum $p^\alpha\simeq E(1;\hat p)$ of the neutrino. This dependence on the direction of propagation arises from the breaking of invariance under rotations. For terrestrial experiments, the direction of the neutrino beam changes as the Earth rotates and so does the coupling with the constant background fields. This change of the hamiltonian with sidereal time will lead to periodic variations on the neutrino oscillation data. In Eq. \refeq{JSD.eq01}, the Lorentz-invariant term as well as the isotropic coefficients in the Lorentz-violating part produce a time-independent part, whereas the remaining terms lead to first and second harmonics of the sidereal phase $\omega_\oplus T_\oplus$,

\begin{eqnarray}\label{JSD.eq02}
(h_\text{eff})_{ab} &=& \frac{m^2_{ab}}{2E}+(\mathcal{C})_{ab}+(\mathcal{A}_s)_{ab}\sin{\omega_\oplus T_\oplus}+(\mathcal{A}_c)_{ab}\cos{\omega_\oplus T_\oplus} \nonumber\\
&&\quad\quad\quad\quad+(\mathcal{B}_s)_{ab}\sin{2\omega_\oplus T_\oplus}+(\mathcal{B}_c)_{ab}\cos{2\omega_\oplus T_\oplus},
\end{eqnarray}
where $T_\oplus$ is the local sidereal time, $\omega_\oplus\simeq2\pi/(23\text{ h }56\text{ min})$ is the sidereal frequency of the Earth, and the amplitudes $(\mathcal{C})_{ab},(\mathcal{A}_{s,c})_{ab}$, and $(\mathcal{B}_{s,c})_{ab}$  are functions of the coefficients $a_L$ and $c_L$.

\section{Applying the theory to experiments}
\label{JSD_Sec3}

The theory described in Sec.\ \ref{JSD_Sec2} can now be applied to different experiments. There are two regimes of this theory that depends on the baseline of the experiment and the energy of the particles studied.

\subsection{Short-baseline experiments}

In the conventional description without Lorentz violation, experiments in which the dimensionless combination $\Delta m^2L/E\ll1$ should not be able to observe oscillations because the oscillation phase is too small. In other words, conventional neutrinos do not have time to oscillate in such a short distance. When this condition is satisfied we can simply drop the first term in Eq. \refeq{JSD.eq02}. Notice that this does not imply that neutrinos do not have mass, it only means that given the energy and baseline, masses cannot be responsible for oscillations.\cite{KM2004c} This is the problem with the observation of oscillations by the Liquid Scintillator Neutrino Detector (LSND) experiment if we assume Lorentz invariance.\cite{LSND2001} If we drop the first term in Eq. \refeq{JSD.eq02}, we still have the possibility of neutrino oscillations produced by the Lorentz-violating part of the hamiltonian. This description has been used by LSND and the Main Injector Neutrino Oscillation Search (MINOS) to look for possible sidereal modulation of their data.\cite{LSND2005,MINOSND2008} Using this description they put the first constraints on coefficients for Lorentz violation in the neutrino sector.\cite{datatables}

\subsection{Long-baseline experiments}

Current and future long-baseline experiments are designed to make precise measurements of the unknown parameters of the conventional model for neutrino oscillations. For this reason, their baseline and neutrino energy satisfy $\Delta m^2L/E\approx1$. In this case the mass term in Eq. \refeq{JSD.eq02} is dominant and the Lorentz-violating part must be treated as a perturbation.\cite{DKM2009} In this perturbative description, the study of constant effects due to Lorentz violation introduced by the second term in Eq. \refeq{JSD.eq02} can be challenging. Nevertheless, the time-dependent contribution introduces a clean sidereal modulation of the oscillation probability over the conventional description. The search for this modulation has been performed recently by MINOS using its far detector.\cite{MINOS:LV2010} Complementary analyses could in the near future be executed using this perturbative description by other long-baseline experiments like ICARUS, K2K, LBNE, NO$\nu$A, OPERA, T2K, and T2KK.

\section{CPT violation and recent neutrino results}
\label{JSD_Sec4}

Conventionally, the search for CPT violation in any sector of the SM is performed by comparisons of the fundamental properties (mass, lifetime, etc.) of a given particle and its corresponding antiparticle. Unfortunately, this method is not consistent with field theory. Since CPT violation implies the breaking of Lorentz symmetry,\cite{greenberg} the study of CPT violation requires a Lorentz-violating framework in which all the fundamental properties of particles and antiparticles are equal as required by the CPT theorem. As we mentioned in Sec. \ref{JSD_Sec1}, CPT violation is already included within the SME. In the case of neutrinos, CPT violation is controlled by $a_L$ in Eq. \refeq{JSD.eq01}. The precise measurements of the mass-squared differences and mixing angles performed during the last decade by different experiments using both neutrinos (accelerator, atmospheric, solar) and antineutrinos (accelerator, atmospheric, reactor) makes tempting the idea of simply comparing these parameters to search for CPT violation. Nonetheless, based on field theory, a CPT-violating quantity must be energy- and momentum-dependent due to the accompanying breaking of Lorentz symmetry. Moreover, terms that break Lorentz symmetry in Eq. \refeq{JSD.eq01} include unconventional energy dependence as well as possible direction dependence; therefore, a complete and systematic study of CPT violation requires the use of a consistent framework. This is precisely what we have described in Sec. \ref{JSD_Sec2}.

Recently, MiniBooNE and MINOS experiments have announced the observation of differences in the way neutrinos and antineutrinos oscillate.\cite{NuResults2010} These results are preliminary; nevertheless, if confirmed as a manifestation of neutrinos and antineutrinos having a different behavior, they would suggest that there exists at least one non-zero coefficient for Lorentz violation.


\end{document}